\documentclass[twocolumn,showpacs,preprintnumbers,amsmath,amssymb,aps]{revtex4}

\usepackage{graphicx}
\usepackage{dcolumn}
\usepackage{bm}

\begin{document}

\title{Transition from Band insulator to Bose-Einstein Condensate superfluid and \\
Mott State of Cold Fermi Gases with Multiband Effects in Optical Lattices}

\author{Ryota Watanabe}
\author{Masatoshi Imada$^1$}%
\affiliation{%
Department of Applied Physics, University of Tokyo, Hongo, Bunkyo-ku, Tokyo 113-8656, Japan \\
$^1$ JST, CREST, Hongo, Bunkyo-ku, Tokyo 113-8656, Japan
}%

\date{\today}

\begin{abstract}
We study two models realized by two-component Fermi gases loaded in optical lattices.
We clarify that multi-band effects inevitably caused by the optical lattices generate a rich structure, when the systems crossover from the region of weakly bound molecular bosons to the region of strongly bound atomic bosons.
Here the crossover can be controlled by attractive fermion interaction.
One of the present models is a case with attractive fermion interaction, where an insulator-superfluid transition takes place.
The transition is characterized as the transition between a band insulator and a Bose-Einstein condensate (BEC) superfluid state.
Differing from the conventional BCS superfluid transition, this transition shows unconventional properties. In contrast to the one-particle excitation gap scaled by the superfluid order parameter in the conventional BCS transition, because of the multi-band effects, a large gap of one-particle density of states is retained all through the transition although the superfluid order grows continuously from zero.
A reentrant transition with lowering temperature is another unconventionality.
The other model is the case with coexisting attractive and repulsive interactions.
Within a mean field treatment, we find a new insulating state, an orbital ordered insulator.
This insulator is one candidate for the Mott insulator of {\it molecular bosons} and is the first example that the orbital internal degrees of freedom of molecular bosons appears explicitly.
Besides the emergence of a new phase, a coexisting phase also appears where superfluidity and an orbital order coexist just by doping holes or particles.
The insulating and superfluid particles show differentiation in momentum space as in the high-$T_{\rm c}$ cuprate superconductors.
\end{abstract}

\maketitle

\section{\label{sec:introduction}Introduction}
Control of interaction strength between particles by utilizing a Feshbach resonance makes it possible to form weakly bound molecules of two Fermi particles with controllable binding energy~\cite{mol1,mol2,mol3,mol4,mol5}.
With the help of this controllability, crossover between the usual BCS superfluid state to the Bose-Einstein condensation (BCS-BEC crossover) was observed in two-component Fermi gases (mixtures of two-hyperfine states)~\cite{crossover1,crossover2}.
In addition to the success in tuning interactions, optical lattices formed by standing waves of light also provide us with ideal systems to study various phenomena, such as the superfluid-Mott insulator transition in Bose systems~\cite{SF-MI}.
This controllability of parameters makes ultracold atomic gases ideal model systems to investigate interacting many-body systems.

Since we can control both interaction strength and lattice potential depth, it is possible to realize any region of energy scales, such as $E_{\rm int}\sim E_{\rm g} \gg E_{\rm kin}$, where $E_{\rm int }$ is the interaction energy, $E_{\rm g}$ is the energy of band gaps and $E_{\rm kin}$ is the kinetic energy.
Thus, we have a chance to study the interpolating region between solid state physics where $E_{\rm g}>E_{\rm int}$ usually stands and molecular physics where $E_{\rm int}\gg E_{\rm g}$ stands.
The intermediate region $E_{\rm int}\sim E_{\rm g}$ is both complicated and interesting.
This is because the interaction mixes two bands separated by the band gap and therefore orbital degrees of freedom play important roles.
In this region, treating molecules as a fixed minimum unit is not justified because several different types of molecular bosons can be formed with different combinations of orbitals.
Internal degrees of freedom of bosons that are often considered in ultracold atomic gases are the spin degrees of freedom of hyperfine states.
It has been theoretically shown that because of spin degrees of freedom of bosons, there can be several nontrivial phases~\cite{spinor1,spinor2,spinor3,spinor4,spinor5}.
However, roles of other types of freedom are rarely considered and remain open questions.
In this paper, we highlight effects of orbital degrees of freedom arising from multi-band effects.

Here, we first review an intriguing experiment with two-component Fermi gases of $^6$Li done by Zwierlein {\it et al.} under the condition with $E_{\rm int}\sim E_{\rm g}$~\cite{experiment-MIT}.
They observed the superfluidity both in the BCS and in the BEC sides with attractive interaction in optical lattices.
We need to pay attention to the density of particles per unit cell $n$.
In this experiment, the density of particles, $n$ is equal to two, which means that there is one particle of each component per unit cell on average.
In this optical lattice, a band gap between the lowest and the second-lowest bands is nonzero under strong periodic lattice potentials.
If the gap is nonzero and there is no interaction, we should obtain a band insulator.
Nevertheless, they observed the superfluidity for the lattice potential strong enough to form a band gap between the lowest and the second-lowest bands.
In addition, they observed the disappearance of the superfluidity by strengthening the lattice potential.
Then, the authors claimed that it was a transition between the superfluidity and the Mott insulator of {\it molecular bosons}.

Considering the experimental condition, however, we conclude that they observed the transition between a band insulator and a superfluid state.
The reasons are the following.
The added attractive interaction was about $E_{\rm int}=7.5 [{\rm kHz}]$ estimated from Ref.~\cite{mol5}, while the critical band gap where the superfluidity vanished was about $E_{\rm g}=37.5 [{\rm kHz}]$ estimated using our model shown in Sec.~\ref{sec:BI-SF}.
The energy cost to add one fermion to the insulator was $E_{\rm g}=37.5 [{\rm kHz}]$ while the energy cost to add a pair of fermions to the insulator was $2E_{\rm g}-E_{\rm int}=67.5 [{\rm kHz}]$.
This means that low-energy excitations are fermionic rather than bosonic ones.
Therefore, it is reasonable to conclude that the transition observed by Zwierlein {\it et al.} is the band insulator-superfluid transition. 

This transition occurs in the case where $E_{\rm int}\sim E_{\rm g}$ stands.
Since this condition is hardly realized in electron systems and is not fully studied, unexpected mechanisms or phenomena as well as internal degrees of freedom of molecular bosons may play important roles.
Some theoretical works suggested possible insulator-superfluid transitions caused by the attractive interactions~\cite{BI-SF1,BI-SF2,BI-SF3,BI-SF4,BI-SF5}.
So far, however, little is known on the character of the transition and on properties of phases around the transition point with thermal effects.
Therefore, detailed analyses on the transition are desired.
How the Mott insulator of {\it molecular bosons} emerges is also an interesting open question.

In this paper, we first focus on a case with attractive interaction between $|\uparrow \rangle$-components and $|\downarrow \rangle$-components (we call two hyperfine states $|\uparrow \rangle$ and $|\downarrow \rangle$), which we discuss in Sec.~\ref{sec:BI-SF}.
This condition is realized by utilizing a Feshbach resonance as in the experiment by Zwierlein {\it et al.}~\cite{experiment-MIT}.
We show that the band insulator-superfluid transition is characterized by a remarkable feature that the superfluid gap is not scaled by the superfluid order parameter and is already large in contrast to the order parameter growing from zero in the vicinity of the transition.
Furthermore, the binding energy of a Cooper pair near the transition point is large enough that a Cooper pair is considered as a molecular boson.
Thus, this transition is characterized by a transition between the band insulator and the Bose-Einstein condensation.
These features were mentioned in Refs.~\cite{BI-SF1,BI-SF5}.
In addition, with decreasing temperatures, a reentrant transition into the non-ordered phase appears.

It is also illuminative to compare this emergence of the superfluidity with a completely different and extreme case of simple one-component bosonic atoms.
In atomic Bose gases with optical lattices, the binding interactions which stabilize bosons, namely Bose atoms are nuclear or electron-nuclei Coulomb interactions, which are much stronger than the lattice potential.
In comparison with the interaction scales in the first system (interacting fermions), the attractive interactions stabilizing atoms are strong enough, where the system shows the superfluidity.
Nevertheless, in Bose systems, there exists an insulator, namely the Mott insulator in optical lattices, if the repulsive interaction between bosons is enhanced.

The emergence of the Mott insulator of {\it molecular bosons} was claimed in previous works~\cite{BI-SF2,BI-SF3}.
They claimed that the Mott insulator emerges from the band insulator by a crossover.
However, it is not a simple problem.
In a Mott state, repulsive interaction is needed to stabilize the insulating phase.
The repulsive interaction between molecular bosons in attractive Fermi systems is caused only when molecules have a spatial extension, which is scaled by the $s$-wave scattering length between opposite components of Fermi gases $a_s$~\cite{m-m} and becomes negligible in the limit of Bose atoms.
This means that the Mott insulator of {\it molecular bosons} does not exist in the strong coupling limit (the limit of Bose atoms) in attractive Fermi systems.
Therefore, the Mott insulator is possible only when the spatial extension or finite interaction range of the molecular boson comes into play.
We then need to introduce repulsive interaction between molecules caused by its spatial extension to find the Mott insulator.
In fact, the Mott insulator of Bose atoms such as two-dimensional $^4$He on the periodic potential of the substrate can be realized through van der Waals repulsive interaction which indeed requires the finite range interaction of extended molecules.

To realize these conditions, we introduce the second model, where one $s$-wave Feshbach resonance between the opposite components and two $p$-wave Feshbach resonances~\cite{p-wave1,p-wave2}, one for the ``$|\uparrow \rangle$-$|\uparrow \rangle$" channel and the other for the ``$|\downarrow \rangle$-$|\downarrow \rangle$" channel, are assumed.
It is easy to realize that the repulsive interactions are mimicking repulsive (van der Waals) interactions in the Bose atom limit.
Such a system may be realized in experiments, though it is difficult so far because of large losses of atoms around resonance points.
In addition to Feshbach resonances induced by magnetic fields which are widely used in experiments, optical Feshbach resonances~\cite{opticalFR1, opticalFR2, opticalFR3, opticalFR4, opticalFR5} and Feshbach resonances induced by dc electric fields~\cite{EFR1, EFR2, EFR3, EFR4} are, in principle, available simultaneously.
With the help of combinations of three different Feshbach resonances, it is possible to tune several interactions.

We show in Sec.~\ref{sec:MI-MB} that one candidate for the Mott insulator of {\it molecular bosons} which is associated with the boson Mott insulator emerges in this system in the form of an orbital ordered insulator (OOI) in the mean-field theory.
The OOI is caused by the orbital internal degrees of freedom of molecular bosons.
If we assume that all the molecular bosons are the same kind, these typical internal degrees of freedom do not exist.
We also investigate the case where the density of particles deviates from $n=2$ and find a new phase where superfluidity and orbital order coexist also emerges.
In this phase, the insulating and the superfluid particles show a sharp differentiation in momentum space.

The organization of this paper is the following.
In Sec.~\ref{sec:BI-SF}, we focus on the case with attractive interaction.
Setting the density of particles per site as $n=2$, we clarify the origin of the band insulator-superfluid transition and reveal the characteristic properties.
In Sec.~\ref{sec:MI-MB}, we discuss the case where there is a repulsive interaction between the same components of fermions coexisting with an attractive interaction between the same components.
If the attractive interaction between the opposite component becomes strong, this system is categorized as a Bose gas with repulsive interaction.
In Sec.~\ref{sec:sum}, we summarize the results and discuss future problems.

\section{\label{sec:BI-SF}Band insulator-BEC Superfluid transition}
\subsection{\label{subsec:model}Model and Mean Field Approximation}
Let us consider two-component Fermi gases loaded in optical lattices.
Both components are assumed to have the same mass $m$ and the interaction between the opposite components is tuned via the Feshbach resonance.
Since a lattice potential is induced by standing waves of light, the 3D lattice potential with the simple cubic symmetry has the form  
\begin{align}
V({\bm r}) =& S_{\rm lattice} E_{\rm r} \Bigl( \sin^2(k_{\rm L} x) +\sin^2(k_{\rm L} y)+\sin^2(k_{\rm L} z) \Bigr),
\end{align}
where $k_{\rm L}$ is a wave number of light, $E_{\rm r}$ is the recoil energy defined as $\hbar ^2 {k_{\rm L}}^2/2m$, and the coefficient $S_{\rm lattice}$ is a constant which can be controlled by tuning the intensity of light.
These conditions lead to the Hamiltonian 
\begin{align}
H =& \sum _{\sigma} \int d {\bm r} f^{\dagger} _{\sigma }({\bm r})\Bigl( -\frac{\nabla ^2}{2m} - \mu + V({\bm r}) \Bigr) f_{\sigma }({\bm r}) \nonumber\\
&-U \int d {\bm r} f ^{\dagger} _{\uparrow }({\bm r}) f ^{\dagger} _{\downarrow }({\bm r}) f _{\downarrow }({\bm r}) f _{\uparrow }({\bm r}),
\label{eq:fullH}
\end{align}
where $f^{\dagger}_{\sigma}$ and $f_{\sigma}$ are annihilation and creation operators of fermions, respectively, $\mu$ is the chemical potential, and $U$ is the coupling constant.
We first transform the Hamiltonian into momentum space by the Fourier transformation, such that
\begin{align}
f_{\sigma }({\bm r}) = &\frac{1}{\sqrt{\Omega}}\sum_{\bf k} f_{{\bf k}\sigma } {\rm e}^{-i{\bf k} \cdot {\bf r}}.
\label{eq:FT}
\end{align}
Substituting Eq.(\ref{eq:FT}) into the kinetic term of the Hamiltonian yields
\begin{align}
H_{\rm kin}= &\sum_{{\bf k}\sigma } \Bigl( \frac{k^2}{2m}-\mu \Bigr) f^{\dagger}_{{\bf k} \sigma }f_{{\bf k} \sigma}
+\sum_{\bf kq} f^{\dagger} _{{\bf k}\sigma } V_{\bf q} f_{{\bf k+q}\sigma }.
\end{align}
Here $V_{\bm q}$ has a finite value only when one of $|q_i|$ equals to $g$, the length of the unit reciprocal vector, and the others vanish.
By this transformation, we find the interaction term of the Hamiltonian 
\begin{align}
H_{\rm int}=& -\frac{U}{\Omega} \sum_{\bf kk^{\prime}q} f^{\dagger} _{{\bf k}+{\bf q}/2 \uparrow } f^{\dagger} _{-{\bf k}+{\bf q}/2 \downarrow } f_{-{\bf k^{\prime }}+{\bf q}/2 \downarrow } f _{{\bf k^{\prime}}+{\bf q}/2 \uparrow}.
\label{eq:fullint}
\end{align}
To focus on the superfluid state, we pick up the term which has zero total momentum for the Cooper pairs, namely 
\begin{align}
H_{\rm int}=& -\frac{U}{\Omega} \sum_{\bf kk^{\prime}} f^{\dagger} _{{\bf k}\uparrow } f^{\dagger} _{{\bf -k}\downarrow } f_{{\bf -k^{\prime }} \downarrow } f _{{\bf k^{\prime}}\uparrow } 
\label{eq:BCSint}
\end{align}
and neglect the others. 
Diagonalizing the one-body parts by the unitary transformation leads to the lattice Hamiltonian 
\begin{align}
H=& \sum _{i{\bf k}\in {\rm 1stBZ}} (\varepsilon ^i_{\bf k} -\mu )c^{i \dagger} _{\bf k} c^i _{\bf k}
-\frac{U}{\Omega} \sum_{ij{\bf kk^{\prime}}} 
c^{i \dagger} _{{\bf k}\uparrow } c^{i \dagger} _{{\bf -k}\downarrow } c^j _{{\bf -k^{\prime }} \downarrow } c^j _{{\bf k^{\prime}}\uparrow }, 
\label{eq:hlattice}
\end{align}
where the superscript $i$ represents the band index and $\varepsilon ^i_{\bf k}$ expresses the band dispersion energy, which is a function of the lattice potential depth $S_{\rm lattice}$.
Differing from the single-band Hubbard model, this lattice Hamiltonian contains more than one band because we consider not only the weak coupling but also the strong coupling regions, where the energy scale of the interaction strength is larger than that of the band gap.
In this Hamiltonian, there is no paring interaction between particles in different bands because we only pick up the BCS interaction term Eq.~(\ref{eq:BCSint}).
For mathematical convenience, we rewrite $U/{\Omega}$ as $U/{\Omega} \rightarrow U/N_{\rm s}$, where $N_{\rm s}$ is the number of lattice sites.
Though the volume $\Omega$ equals to $N_{\rm s} N_{\rm b}$, where $N_{\rm b}$ is the number of considered bands, we set $U/N_{\rm b} \rightarrow U$ to follow the conventional notation.

To discuss the band insulator-superfluid transition induced by the attractive interaction, we use the extended BCS mean field approximation introduced by Leggett~\cite{Leggett}.
We first introduce a superfluid order parameter 
\begin{align}
\Delta =& \frac{U}{N_{\rm s}} \sum_{i{\bf k}} \langle c^i_{{\bf -k}\downarrow} c^i_{{\bf k}\uparrow} \rangle
\label{eq:sf-order}
\end{align}
as in the conventional BCS mean field approximation.
Using the mean field Eq.~(\ref{eq:sf-order}), we find a mean field BCS Hamiltonian
\begin{align}
H_{\rm MF} =& \sum_{i{\bf k}} 
\Biggl( \left[ 
\begin{array}{cc}
c^{i \dagger}_{k \uparrow } & c^i _{-k \downarrow } \\
\end{array} 
\right]
\left[ 
\begin{array}{cc}
\varepsilon ^i _{\bf k} -\mu & - \Delta \\
-\Delta^{\dagger} & -\varepsilon ^i _{\bf k} +\mu \\
\end{array} 
\right] 
\left[ 
\begin{array}{c}
c^i _{k \uparrow } \\
c^{i \dagger}_{-k \downarrow }  \\
\end{array} 
\right] \nonumber\\
&+ \varepsilon ^i _{\bf k} -\mu \Biggr)
+ \frac{N_{\rm s}|\Delta|^2}{U} .
\label{eq:MFH}
\end{align}
To diagonalize this Hamiltonian, we use the Bogoliubov transformation:
\begin{align}
c^i _{{\bf k}\uparrow}=& u^i _{\bf k} \alpha^i _{{\bf k}\uparrow}- v^i _{\bf k} \alpha^{i \dagger} _{{\bf -k}\downarrow} \label{eq:Bog1}\\
c^{i \dagger} _{{\bf -k}\downarrow}=& u^i _{\bf k} \alpha^{i\dagger} _{{\bf -k}\downarrow}+ v^{i \ast} _{\bf k} \alpha^{i} _{{\bf k}\uparrow},
\label{eq:Bog2}
\end{align}
where $\alpha^i _{{\bf k}\sigma}$ and $\alpha^{i \dagger} _{{\bf k}\sigma}$ is annihilation and creation operators of quasiparticles, respectively.
We need to determine the coefficients $u^i_{\bf k}$ and $v^i_{\bf k}$ in order to diagonalize the mean field Hamiltonian.
As in the conventional BCS theory, we find
\begin{align}
{u^i _{\bf k}}^2 =& \frac{1}{2} \Bigl( 1+ \frac{\xi^i_{\bf k}}{E^i _{\bf k}}\Bigr) \\
{v^i _{\bf k}}^2 =& \frac{1}{2} \Bigl( 1- \frac{\xi^i_{\bf k}}{E^i _{\bf k}}\Bigr),
\end{align}
where 
\begin{align}
\xi^i_{\bf k}=&\varepsilon ^i_{\bf k}-\mu \\
E^i_{\bf k}=&\sqrt{{\xi^i_{\bf k}}^2+\Delta^2}.
\label{eq:bogeigen}
\end{align}
Without loss of generality, we may set both $u^i_{\bf k}$ and $v^i_{\bf k}$ real.
By this Bogoliubov transformation, we find diagonalized Hamiltonian
\begin{align}
H=& \sum_{i{\bf k}} (\xi^i_{\bf k}-E^i_{\bf k} \alpha^i_{{\bf k}\uparrow}\alpha^{i \dagger}_{{\bf k}\uparrow}+
E^i_{\bf k} \alpha^{i \dagger}_{{\bf k}\downarrow}\alpha^i_{{\bf k}\downarrow})+ \frac{N_{\rm s}|\Delta|^2}{U} .
\end{align}
From this Hamiltonian, the free energy is obtained as 
\begin{align}
F-\mu N =& \sum_{i{\bf k}}(\xi ^i _{\bf k}-E^i_{\bf k}) +\frac{N_{\rm s}|\Delta|^2}{U} \nonumber\\ 
&-2k_{\rm B}T \sum_{i{\bf k}} {\rm ln} \bigl[1+e^{-E^i_{\bf k}/k_{\rm B}T} \bigr],
\label{eq:free}
\end{align}
where $k_{\rm B}$ represents the Boltzmann constant.
The self-consistency condition on the mean field order parameter $\Delta $ defined in Eq.~(\ref{eq:sf-order}) yields the self-consistent equation or the gap equation 
\begin{align}
\Delta =& \frac{U}{N_{\rm s}} \sum_{i{\bf k}} \langle c^i _{{\bf -k}\downarrow} c^i _{{\bf k} \uparrow }\rangle = \frac{U}{N_{\rm s}}\sum_{i{\bf k}} \frac{\Delta}{2 E^i_{\bf k}}\bigl( 1-2f(E^i_{\bf k})\bigr), 
\label{eq:gap}
\end{align}
where $f(\varepsilon )$ is the Fermi distribution function.
We also need another self-consistent equation to determine the chemical potential $\mu$ because in the  larger $U$ region, contrary to the smaller $U$ region, the chemical potential largely deviates from the Fermi energy.
Using the thermodynamic relation, we find
\begin{align}
N=& -\frac{\partial G}{\partial \mu},
\end{align}
where $G=F-\mu N$ and $N$ is the total number of particles.
From this relation, we find the equation 
\begin{align}
N=& \sum_{i{\bf k}} \Bigl( 1-\frac{\xi^i_{\bf k}}{E^i_{\bf k}}\bigl( 1-2f(E^i_{\bf k})\bigr)\Bigr),
\label{eq:number}
\end{align}
which determines the chemical potential. 

With this mean field approximation, given $U$ and $N$ as input parameters, $\Delta$ and $\mu$ are determined self-consistently from Eqs.~(\ref{eq:gap}) and (\ref{eq:number}). 

\subsection{\label{subsec:nresults}Numerical Results}
We show some of the key results of the above mean field approximation for the cubic lattice in the following.
The superfluid transition induced by the attractive interaction has some outstanding properties that differ from those of the conventional BCS superfluid transition.
Unconventional transitions of the one-particle density of states (DOS) and reentrant behavior should be observed around the transition.

\begin{figure}
\includegraphics[width=6cm]{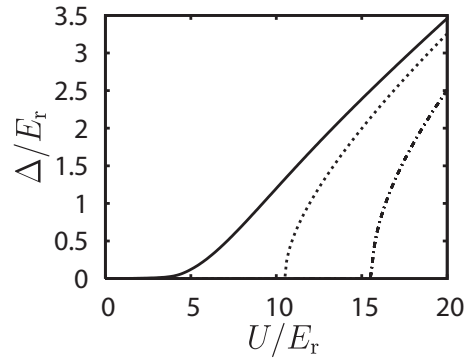}
\caption{\label{fig:delta} Superfluid order parameter $\Delta$ as a function of $U$ at $T=0$.
The solid, dotted and dash-dotted lines correspond to the result for the lattice potential $S_{\rm lattice}=0$, $S_{\rm lattice}=5$ and $S_{\rm lattice}=10$, respectively.
Quantum phase transitions to superfluid phase are found at nonzero $U$ for $S_{\rm lattice}=5$ and $S_{\rm lattice}=10$.
}
\end{figure}

In Fig.~\ref{fig:delta}, we plot the superfluid order parameter as a function of the interaction strength $U$ with $n=2$ at absolute zero temperature.
While, in the absence of the lattice potential, the superfluid phase dominates over $U>0$ and the transition point is at $U=0$, the insulator-superfluid transition occurs at a nonzero $U$ with a finite band gap between the lowest and the second-lowest bands in the presence of the lattice potential.
By assuming $|\Delta|/|\xi^i_{\bf k}| \ll 1$, expansion of $E^i_{\bf k}$ in the free energy yields 
\begin{align}
F=& \Bigl( \frac{N_{\rm s}}{U}-\sum_{{\bf k}i}\frac{|\xi^i_{\bf k}|}{2{\xi^i_{\bf k}}^2} \Bigr) |\Delta|^2+ \sum_{{\bf k}i} \frac{|\xi^i_{\bf k}|}{8 {\xi^i_{\bf k}}^4} |\Delta|^4 +O(|\Delta|^6) .
\end{align}
Here, we drop the constant term.
The coefficient of the quartic term is definitely real and positive, while the quadratic term changes the sign as a function of $U$.
This is a typical feature of the second-order transition in the Landau expansion, which is consistent with the behavior of the order parameter shown in Fig.~\ref{fig:delta}.
This expansion is justified because the chemical potential is located in the band gap near the transition point, which means that all $\xi^i_{\bf k}$ are not equal to zero.

As in Ref.~\cite{BI-SF1}, the critical value of interaction $U_{\rm c}$ is roughly estimated as follows: Let us consider the energy required to excite a pair of fermions from the lower to the higher band. 
If there is no interaction between the particles, the energy is equal to twice the energy of the band gap, namely $2E_{\rm g}$. 
However, with the interaction, the energy is modified to about $2E_{\rm g}-U$. 
This is because the pair in the second-lowest band gains $-U$ from the BCS channel which is prohibited by Pauli principle in the band insulator. 
At sufficiently large $U$, some configurations lower the energy than that of the band insulator. 
From this insight, we conclude that the pair formation in the higher bands causes this transition and that
the transition point $U_{\rm c}$ is roughly determined as $U_{\rm c} \sim  2E_{\rm g}$.

\subsubsection{\label{ssubsec:DOS}One-Particle Density of States (DOS)}
Since both the band insulating and the superfluid states should show gapped DOS, it is worth investigating the transition of DOS over the transition.
In our calculation, DOS originating from particles in the $i$th-band $D^i(\omega)$ is obtained by using the relation 
\begin{align}
D^i(\omega)=& -\frac{1}{\pi} {\rm sgn} \omega \sum_{\bf k} {\rm Im} G^{ii}({\bm k},\omega) , 
\label{eq:DOS}
\end{align}
where $G^{ii}({\bm k}, \omega)$ is the Fourier component of the one-particle Green's function 
\begin{align}
G^{ii}({\bm k}, t)=& -i \langle T_t (c^i_{{\bf k}\uparrow}(t) c^{i\dagger}_{{\bf k}\uparrow}(0) )\rangle . 
\end{align}
Here, $T_t$ is the time-ordering operator.
Since DOS originating from the down-spin component is the same as that of the up-spin component in this case, we define DOS $D^i(\omega)$ only by the Green's function of the up-spin component.
Using the Bogoliubov transformation defined in Eqs.~(\ref{eq:Bog1}) and (\ref{eq:Bog2}), Green's function reads 
\begin{align}
G^{ii}({\bm k}, t)=& -i {u^i_{\bf k}}^2\langle T_t (\alpha^i_{{\bf k}\uparrow}(t) \alpha^{i\dagger}_{{\bf k}\uparrow}(0) )\rangle \nonumber\\
&-i{v^i_{\bf k}}^2\langle T_t (\alpha^{i\dagger}_{-{\bf k}\downarrow}(t) \alpha^i_{-{\bf k}\downarrow}(0) )\rangle. 
\end{align}
The Fourier transformation yields 
\begin{align}
G^{ii}({\bm k}, \omega)=& \frac{{u^i_{\bf k}}^2}{\omega-E^i_{\bf k}+i\delta}+\frac{{v^i_{\bf k}}^2}{\omega+E^i_{\bf k}-i\delta}. 
\label{eq:Gkw}
\end{align}
Substituting Eq.~(\ref{eq:Gkw}) into Eq.~(\ref{eq:DOS}), we find 
\begin{align}
D^i(\omega)=& \sum_{\bf k} \Bigl({u^i_{\bf k}}^2 \delta(\omega-E^i_{\bf k})+{v^i_{\bf k}}^2 \delta(\omega+E^i_{\bf k}) \Bigr) .
\label{eq:DOS-2}
\end{align}

\begin{figure}
\begin{center}
\begin{tabular}{c}
\includegraphics[width=6cm]{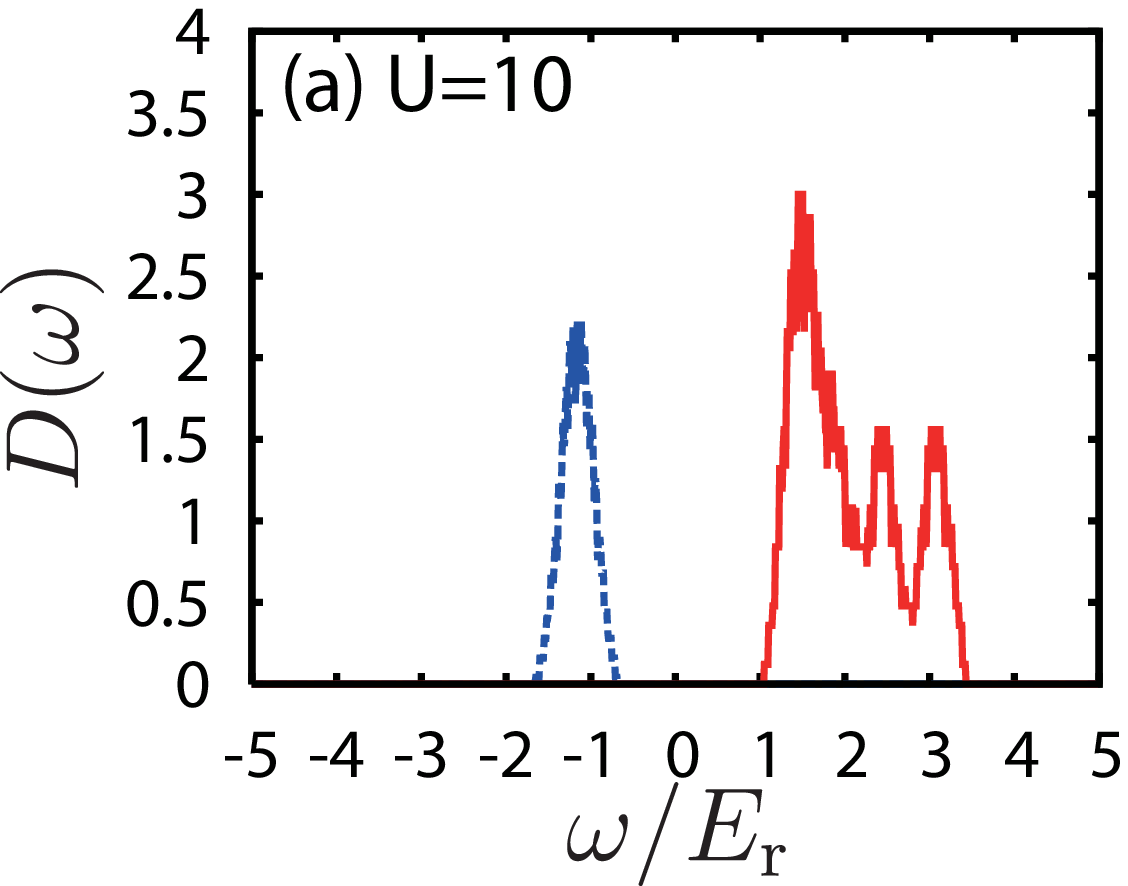} \\
\includegraphics[width=6cm]{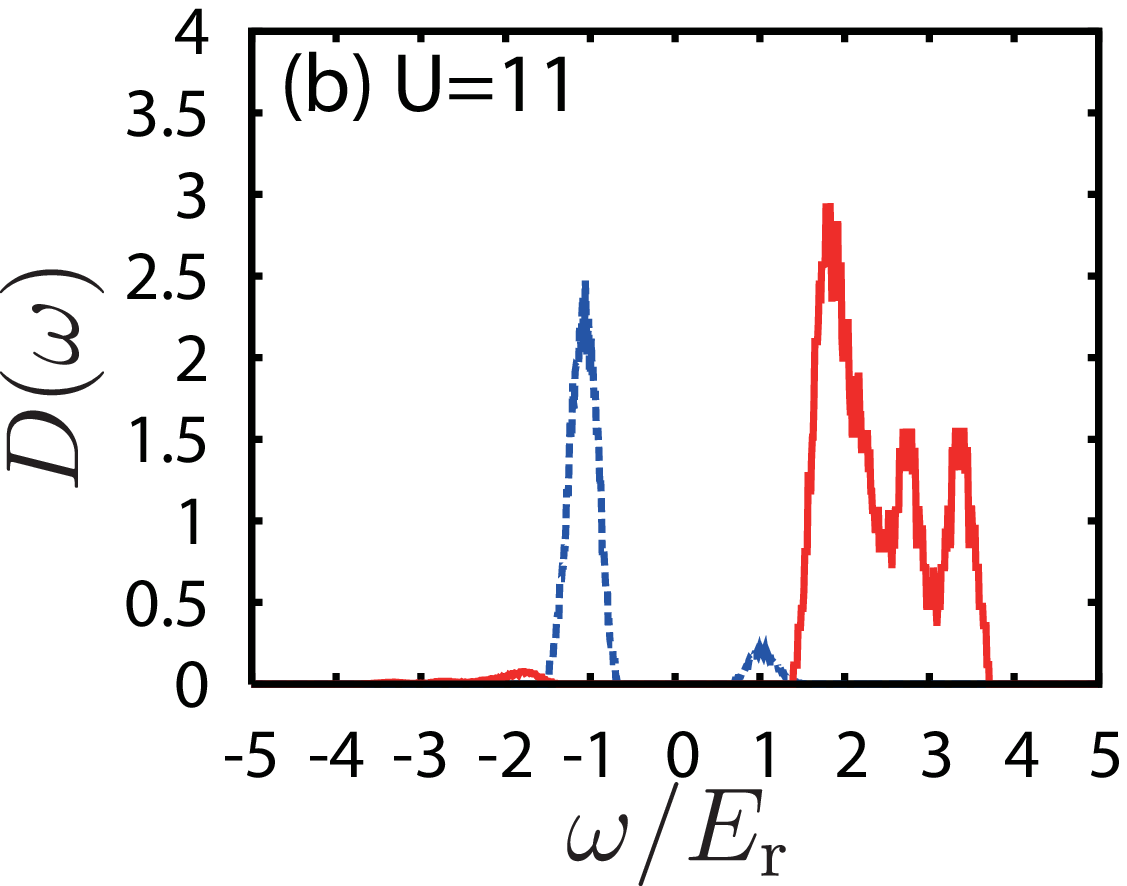} \\
\includegraphics[width=6cm]{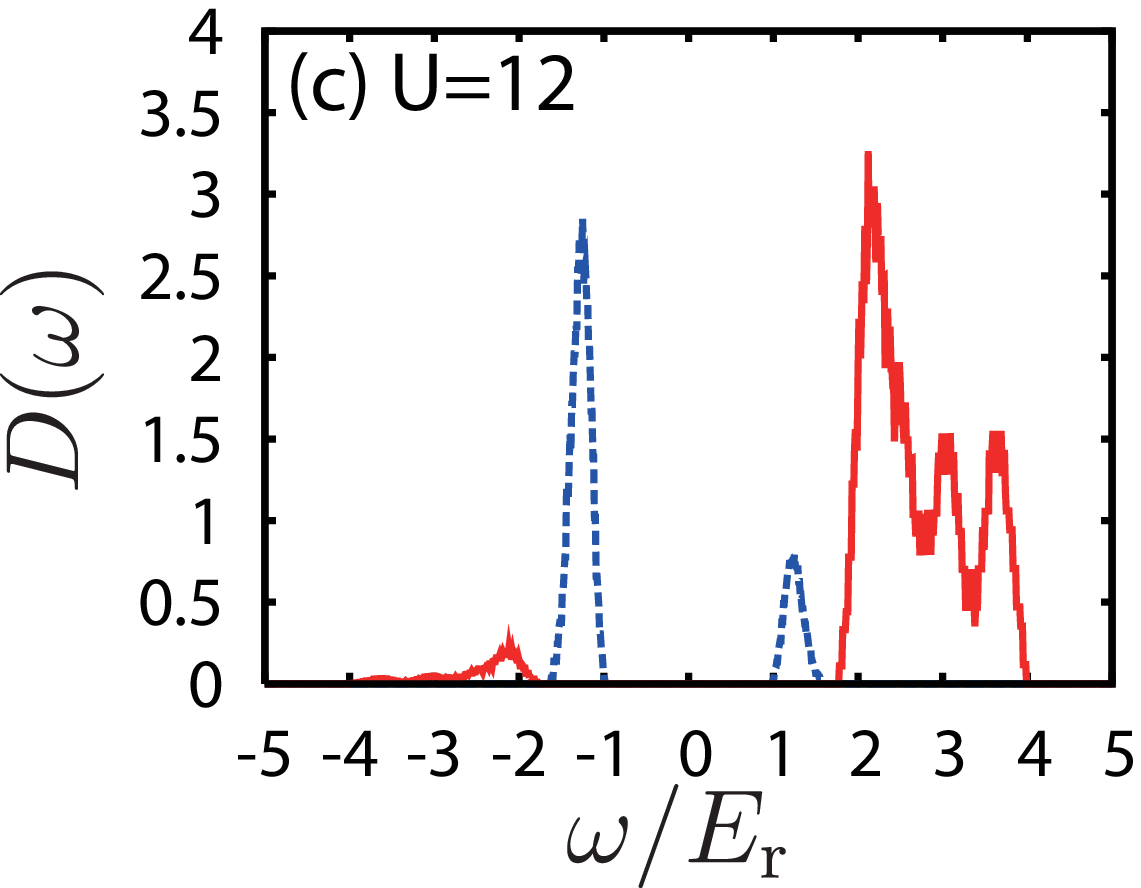} \\
\includegraphics[width=6cm]{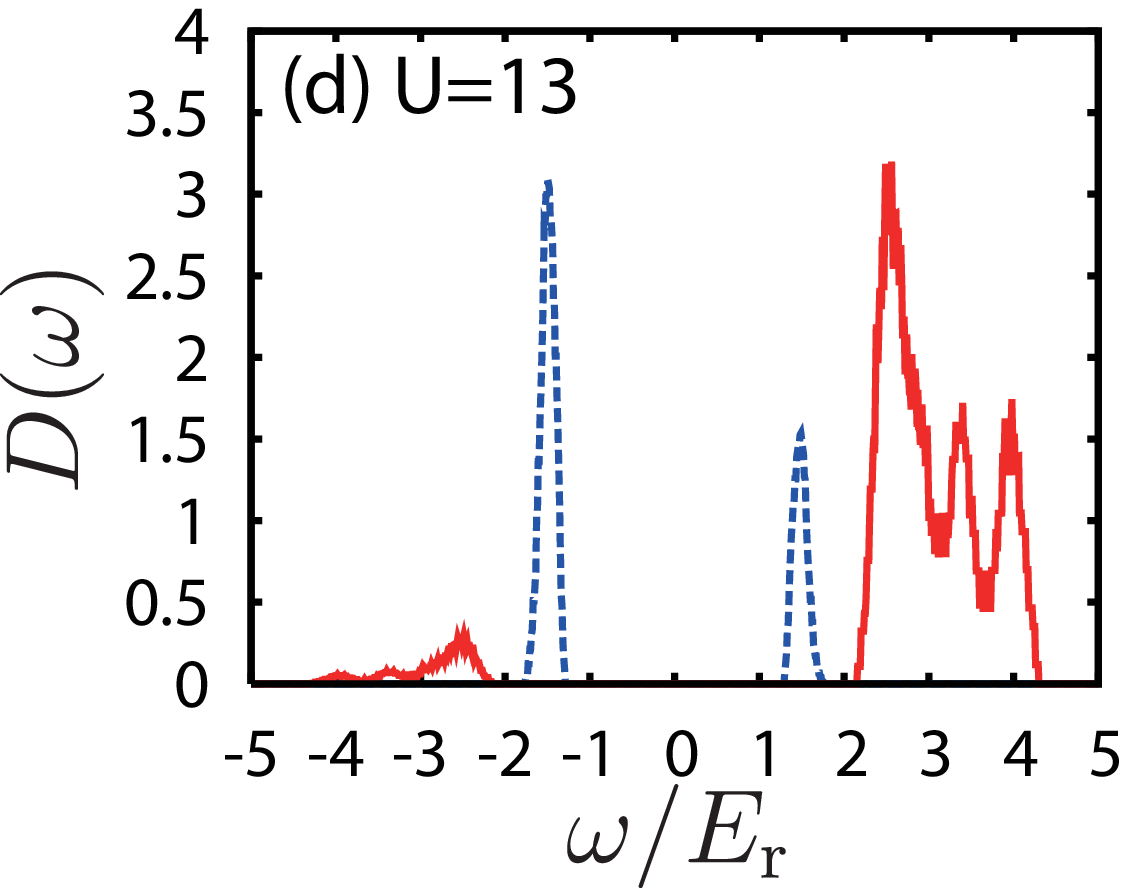} \\
\end{tabular}
\end{center}
\caption{\label{fig:DOS} (color online) One-particle density of states (DOS) $D(\omega)$ at $S_{\rm lattice}=5$.
The dashed (blue) line shows the lowest-band component $D^1(\omega)$ while the solid (red) one shows that of the second-lowest three bands $\tilde {D^2}(\omega)=\sum_{i=2,3,4} D^i(\omega)$.
The result at $U=10$ is in the band insulating phase and those at $U=11,12,13$ are in the superfluid phase.
Both in the band insulator and in the superfluid, DOS stays always zero around the chemical potential $\omega=0$ through the transition.
}
\end{figure}
Let us call the lowest energy band ``1" with the $1s$ symmetry and the three-folded second-lowest bands ``2", ``3" and ``4" with the $2p$ symmetry.
In Fig.~\ref{fig:DOS}, we show the one-particle DOS $D^1(\omega)$ and $\tilde {D^2}(\omega)= \sum_{i=2,3,4} D^i(\omega)$ near the transition point by using Eq.~(\ref{eq:DOS-2}), where the superscript denotes the band index.
In the conventional BCS superfluid states, the excitation gap scales with the superfluid order parameter and we should find the coherence peak of DOS around the chemical potential.
In Fig.~\ref{fig:DOS}, however, no visible change is seen around the chemical potential.
Instead, for example, at $U=11$, we find a growth of the DOS $D^1(\omega)$ around $\omega/E_{\rm r}\sim 1$ and that of the DOS $\tilde {D^2}(\omega)$ around $\omega/E_{\rm r}\sim -2$.
This is indeed the evidence for the band insulator-superfluid transition.
The growing DOS $\tilde {D^2}(\omega)$ below the chemical potential corresponds to the bound states of the pair of particles in the higher bands and the growing DOS $D^1(\omega)$ above the chemical potential corresponds to the bound states of the pair of holes in the lower band.
Therefore, strongly bound Cooper pairs are formed and condense near the transition point.
In other words, the insulator-superfluid transition induced by the attractive interaction is that between the band insulator and the BEC superfluid.
This result also indicates that the energy needed to excite one quasi-particle or to break up a Cooper pair is as large as that of the band gap $E_{\rm g}$ even at the transition point.
This means that nonzero excitation gap should exist on the transition point, which is consistent with previous results~\cite{BI-SF1,BI-SF5}.
Therefore, in the superfluid state near the transition point, the superfluid order parameter $\Delta$ is much smaller than the superfluid gap $\Delta_{\rm gap}$.
This is in marked contrast with the conventional superfluid state, where the superfluid gap scales with the amplitude of the superfluid order parameter.
Photoemission spectroscopy on cold atomic gases recently developed~\cite{PE} would reveal this unconventional transition of DOS and give us full understanding of the band insulator-BEC superfluid transition induced by the attractive interaction.

\subsubsection{\label{ssubsec:finiteT}Finite Temperature : Reentrant Transition}
\begin{figure}
\includegraphics[width=6cm]{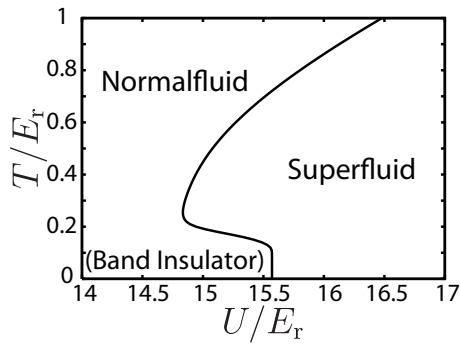}
\caption{\label{fig:PD} Phase diagram in the plane of temperature $T$ and interaction strength $U$ at $S_{\rm lattice}=10$.}
\end{figure}
\begin{figure}
\begin{center}
\begin{tabular}{c}
\includegraphics[width=6cm]{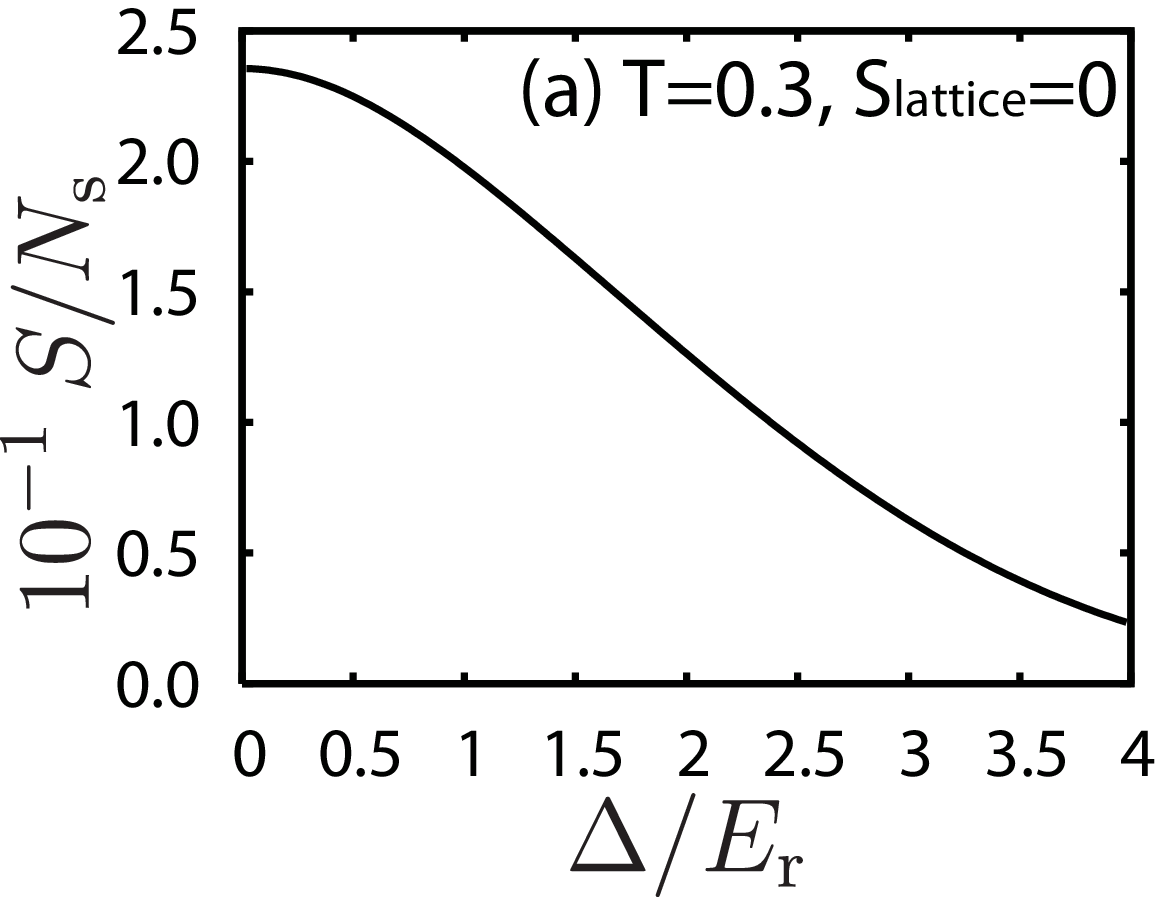} \\
\includegraphics[width=6cm]{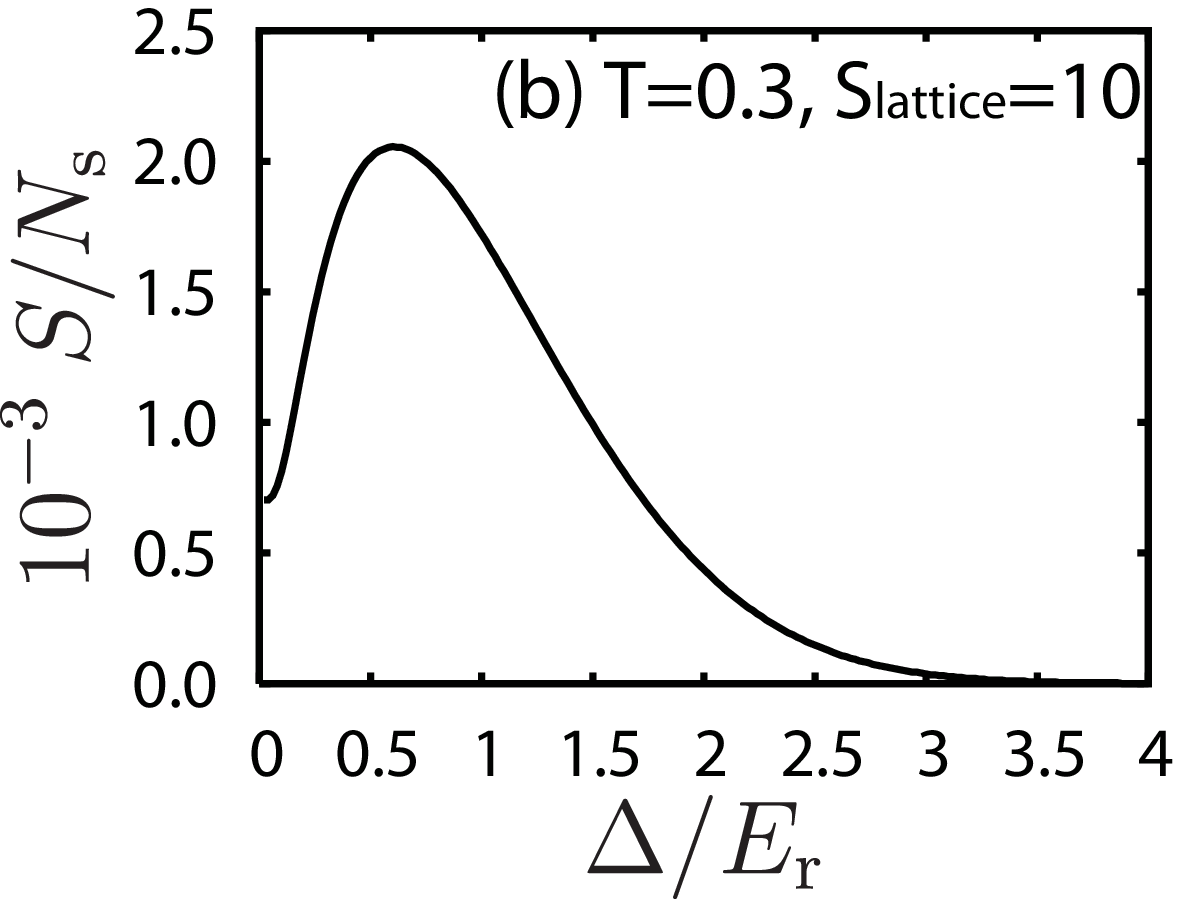} \\
\includegraphics[width=6cm]{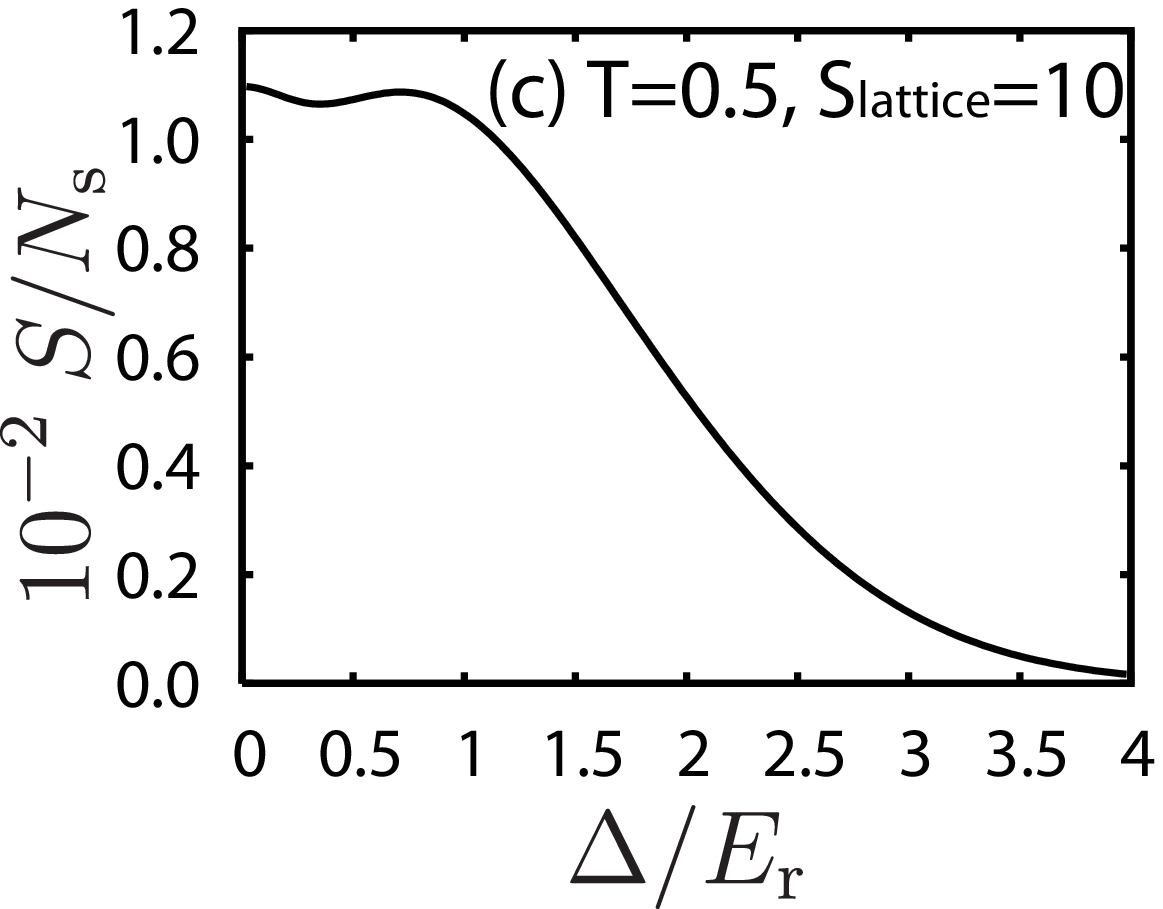} \\
\includegraphics[width=6cm]{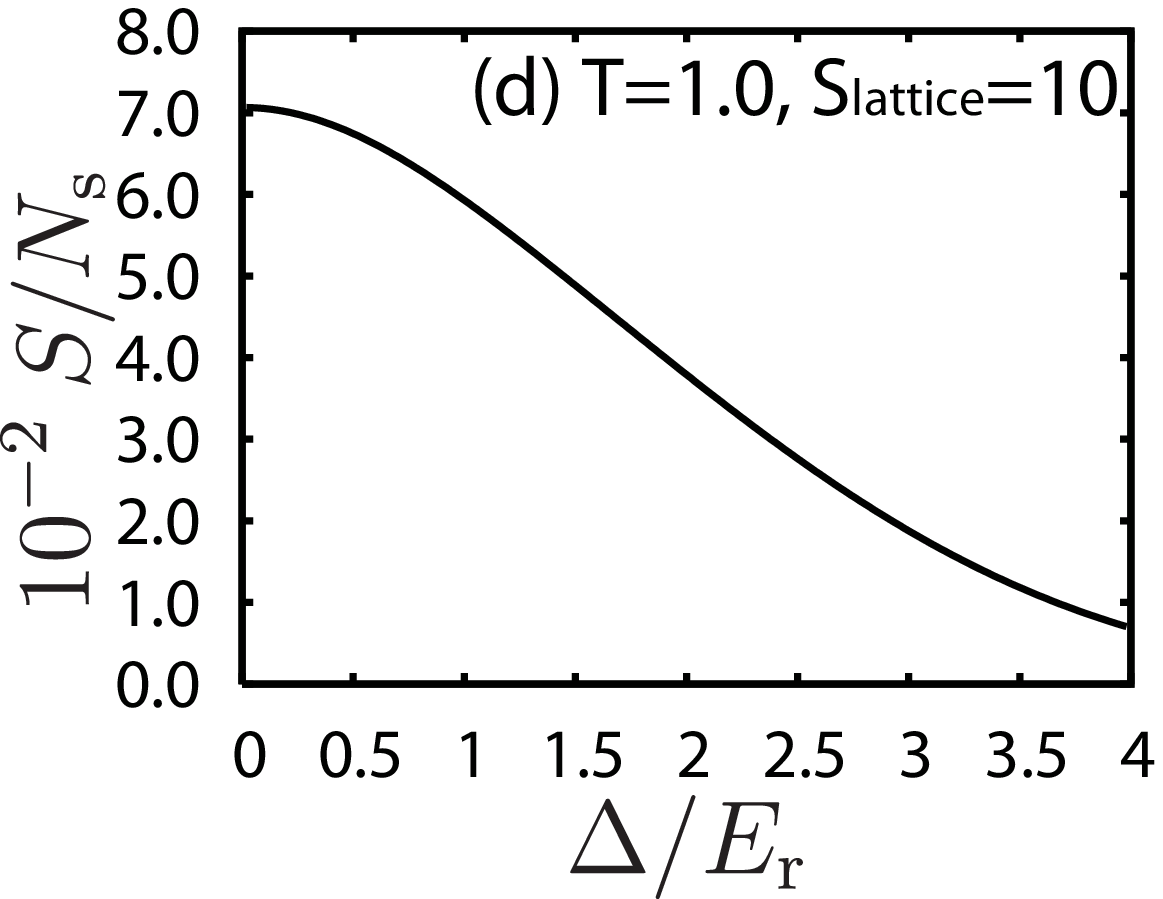} \\
\end{tabular}
\end{center}
\caption{\label{fig:entropy} Entropy as a function of $\Delta$ for several choices of lattice potentials and temperatures.
We take $k_{\rm B}=1$ in this calculation, which means that temperature has the same unit as energy and entropy is dimensionless. 
At a nonzero lattice potential and at sufficiently low temperatures, the entropy has a peak at nonzero $\Delta$, while without the potential, it monotonically decays with increase in $\Delta$.}
\end{figure}

For experimental observations, thermal effects have to be clarified because it is still difficult to cool trapped atoms down to $T \sim 0.01T_{\rm F}$, where $T_{\rm F}$ is the Fermi temperature.
We now discuss finite temperature effects around the band insulator-BEC superfluid transition point.
In Fig.~\ref{fig:PD}, we show the phase diagram at $S_{\rm lattice}=10$ in 3D with the simple cubic symmetry obtained by the mean field approximation.
An outstanding point of the phase diagram is the reentrance of the non-ordered phase, where the lowest critical value of $U_{\rm c}\sim 14.8$ is realized at $T=0.25E_{\rm r}$. 

To reveal the mechanism of this reentrance, we calculate the entropy $S$ as a function of the order $\Delta$.
The equilibrium state is realized at the minimum point of the free energy $F=E-TS$ at finite temperatures, where $E$ is the internal energy of the system.
Therefore, the system takes the phase with the maximum entropy when temperature gets high.
At higher temperatures, the system usually shows a non-ordered phase because the entropy of an ordered phase is usually smaller than that of a non-ordered phase.
However, if the entropy of an ordered phase is larger than that of a non-ordered phase, we observe an ordered phase at higher temperatures.

Using the thermodynamic relation 
\begin{align}
S=&-\frac{\partial F}{\partial T}, 
\end{align}
we find 
\begin{align}
ST=& \sum_{{\bf k}i} \Bigl[ 2k_{\rm B}T {\rm ln} \bigl[ 1+e^{-E^i_{\bf k}/k_{\rm B}T}\bigr] +2 E^i_{\bf k} f(E^i_{\bf k}) \Bigr]
\end{align}
from Eq.~(\ref{eq:free}).
Using this form, we plot the entropy as a function of $\Delta$ in Fig.~\ref{fig:entropy}.
Without a lattice potential, the entropy takes its maximum always at $\Delta=0$ irrespective of temperatures.
On the other hand, with sufficiently strong lattice potential and at low temperatures, the entropy takes its maximum at a nonzero value of $\Delta$.
This smaller value of the entropy at $\Delta=0$ is caused by the fact that the non-ordered state at $\Delta=0$ is nearly a band insulator.
On the other hand, when the temperature exceeds the band gap, entropy takes its maximum at $\Delta=0$.
Thus, at relatively low temperatures, we observe the ordered (superfluid) phase in higher temperature region whereas, at relatively high temperatures, we observe the ordered (superfluid) phase in lower temperature region.
These are the reasons why we observe the reentrance of the non-ordered phase.

The larger entropy in the superfluid phase is easy to understand when we consider detailed band structures.
Comparing the excitation gap in Fig.~\ref{fig:DOS}(a) with that in Fig.~\ref{fig:DOS}(b), we find that the gap in the superfluid phase is smaller than that in the band insulating phase.
Therefore, at nonzero temperatures, thermal excitations make the larger entropy in the superfluid phase than that in the band insulating phase.

As mentioned in Ref.~\cite{BI-SF5}, the smaller superfluid gap is caused by particle-hole asymmetry of our model.
In fact in Ref.~\cite{BI-SF1}, this reentrant behavior was not clear because the toy model of Ref.~\cite{BI-SF1} has particle-hole symmetry.
Considering the difference between the models, we conclude that larger DOS above the chemical potential than that below the chemical potential helps the clear emergence of the reentrance because of its large deviation from particle-hole symmetry.
If the degeneracy of the second-lowest bands increases, experimental observations of the reentrance becomes easier.

Although this calculation is based on the mean field approximation, strong fluctuations, especially phase fluctuations need to be considered around the BEC transition.
When we consider phase fluctuations, the entropy in the superfluid phase becomes even larger, which means that the reentrance becomes clearer.

The phase diagram which we obtained suggests that any band insulator at $T=0$ may undergo a transition into superfluid or superconducting states by thermal excitations.
The necessary condition is only a reasonable value of attractive interaction between particles though it is difficult to realize in other systems than the cold atom system.

\section{\label{sec:MI-MB}Mott state of molecular bosons}
In attractive Fermi systems, we found the band insulating state in weak coupling regions and the superfluid state in strong coupling regions as in previous section.
We can consider the strong coupling regions of attractive Fermi systems as Bose systems.
Although, in Bose systems, there is a Mott insulating phase if repulsive interaction between atoms is enhanced, we could find no Mott state in strong coupling regions of attractive Fermi systems.
This is because the repulsive interaction between molecular bosons is scaled by the $s$-wave scattering length of $|\uparrow \rangle$-$|\downarrow \rangle$ channel and goes to zero in the strong coupling limit.
However, in Bose systems, repulsive interactions stabilizing Mott states originate from van der Waals interactions, which are not included in attractive Fermi systems.
Thus, to find Mott states of {\it molecular bosons}, an additional repulsive interaction needs to be introduced.
In the following, we propose a new system which can be realized in experiments and show one candidate for the Mott insulator of {\it molecular bosons}.
The insulator has a strong relation to orbital internal degrees of freedom of molecular bosons.
In addition, we investigate cases where $n$ deviates from 2, and show a new coexisting phase and its typical properties.

\subsection{\label{subsec:model2}Model and Mean Field Approximation}
Since magnetic, electric and optical Feshbach resonances are available simultaneously, several types of interactions may be tuned simultaneously.
In this section, we consider the case where $|\uparrow \rangle$-$|\downarrow \rangle$ channel is controlled by a magnetic Feshbach resonance, and $|\uparrow \rangle$-$|\uparrow \rangle$ and $|\downarrow \rangle$-$|\downarrow \rangle$ channels are controlled by optical Feshbach resonances.
A necessary condition to control all the interactions is that all the three resonance points are well separated.

From now on, we consider the case where $ |\uparrow \rangle$-$|\downarrow \rangle $ interaction has a $s$-wave symmetry and is attractive with the value $-U$, and both $|\uparrow \rangle$-$|\uparrow \rangle$ and $|\downarrow \rangle$-$|\downarrow \rangle$ interactions have $p$-wave symmetries and are repulsive with the same value, $W$.
Since these interactions are originally short ranged and gases are dilute, we can treat them as interactions within same lattice sites in lattice models.
As an approximation, we consider the $p$-wave interactions as homogeneous for simplicity, though they have dependence on scattering angles.
Although this approximation may cause quantitative difference on results, there is no qualitative difference.

We naively find that molecular bosons are formed because of the attractive interaction and that molecules form a repulsively interacting Bose gas.
However, this is not such a simple case as a repulsively interacting Bose gas.
Although treating a bosonic atom as a minimum unit is justified, treating a molecular boson as a simple boson is not justified when the interaction energy between molecules are comparable to the binding energy of a molecular boson.
In addition, more than one energy bands are needed to investigate the repulsively interacting molecular Bose gas because of the Pauli exclusion principle of fermions.
Considering these conditions, we introduce a simplest Hamiltonian, namely a two-dimensional two-orbital Hamiltonian
\begin{align}
H=& \sum_{i{\bf k}\sigma} (\varepsilon^i_{\bf k}-\mu )c^{i\dagger}_{{\bf k}\sigma}c^i_{{\bf k}\sigma} -\frac{U}{N_{\rm s}}\sum_{ij{\bf kk^{\prime}}} c^{i\dagger}_{{\bf k}\uparrow}c^{i\dagger}_{-{\bf k}\downarrow} c^j_{-{\bf k^{\prime}}\downarrow}c^j_{{\bf k^{\prime}}\uparrow}\nonumber\\ &+\frac{W}{N_{\rm s}}\sum_{{\bf kk^{\prime}}\sigma} c^{1\dagger}_{{\bf k}\sigma}c^1_{{\bf k}\sigma}c^{2\dagger}_{{\bf k^{\prime}}\sigma}c^2_{{\bf k^{\prime}}\sigma},
\label{eq:2oH}
\end{align}
where $\varepsilon ^i_{\bf k}$ represents the energy dispersions 
\begin{align}
\varepsilon^1_{\bf k}=&-2t(\cos(k_x)+ \cos(k_y)) +4t^{\prime} \cos(k_x)\cos(k_y) \\
\varepsilon^2_{\bf k}=&-2t(\cos(k_x)+ \cos(k_y)) +4t^{\prime} \cos(k_x)\cos(k_y) \nonumber\\ &+ \Delta E. 
\label{eq:edispersion}
\end{align}
Here, we retain only the Hartree-Fock and BCS terms in the mean field treatment for simplicity and we set the lattice constant unity.
While the energy splitting of the two bands is given by the amplitude of the periodic lattice potential $S_{\rm lattice}$ in Sec.~\ref{sec:BI-SF}, in this section, we here take the energy splitting $\Delta E$ as an parameter in Eq.~(\ref{eq:edispersion}).
The first and the second terms in Eq.~(\ref{eq:2oH}) are the same as in Sec.~\ref{sec:BI-SF}, though energy dispersions are different.
The last term in Eq.~(\ref{eq:2oH}) is caused by $p$-wave Feshbach resonances.
As we mentioned above, interactions are limited local within same sites.
Thus, interactions caused by $p$-wave resonances must be between different orbitals because of Pauli principle.
We take $t=1$ as the energy unit and control the parameters $\Delta E$ and $t^{\prime}$, which determines the band structure for non-interacting particles.
With these band structures, we investigate transitions among superfluid, normalfluid and insulators as a function of the two interactions $U$ and $W$.

\begin{figure}
\begin{center}
\includegraphics[width=6cm]{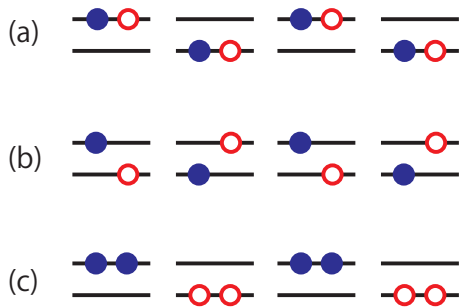} 
\end{center}
\caption{\label{fig:arrange} (color online) Configurations of orders $m_1$, $m_2$, and $m_3$.
(a) is the arrangement of $m_1$, (b) is the arrangement of $m_2$, and (c) is the arrangement of $m_3$.
Filled (blue) particles represent $| \uparrow \rangle$ and open (red) particles are for $| \downarrow \rangle$ components.
}
\end{figure}

To solve the Hamiltonian Eq.~(\ref{eq:2oH}) by using mean field approximations, we introduce mean fields 
\begin{align}
\Delta=& \frac{1}{N_{\rm s}}\sum_{i{\bf k}} \langle c^i_{{\bf -k}\downarrow}c^i_{{\bf k}\uparrow} \rangle \\
n=& \frac{1}{N_{\rm s}} \sum_{{\bf r}\sigma} (n^1_{{\bf r}\sigma}+n^2_{{\bf r}\sigma}) \\
m_1=&\frac{1}{N_{\rm s}} \sum_{\bf r} (n^1_{{\bf r}\uparrow}+n^1_{{\bf r}\downarrow}-n^2_{{\bf r}\uparrow}-n^2_{{\bf r}\downarrow})e^{i{\bf Q}\cdot{\bf r}} \label{eq:WMFn}\\
m_2=&\frac{1}{N_{\rm s}} \sum_{\bf r} (n^1_{{\bf r}\uparrow}+n^2_{{\bf r}\downarrow}-n^1_{{\bf r}\downarrow}-n^2_{{\bf r}\uparrow})e^{i{\bf Q}\cdot{\bf r}} \\
m_3=&\frac{1}{N_{\rm s}} \sum_{\bf r} (n^1_{{\bf r}\uparrow}+n^2_{{\bf r}\uparrow}-n^1_{{\bf r}\downarrow}-n^2_{{\bf r}\downarrow})e^{i{\bf Q}\cdot{\bf r}} ,
\label{eq:WMF3}
\end{align}
where ${\bm Q}=(\pi, \pi)$ and $n^i_{{\bf r}\sigma}= \langle c^{i\dagger}_{{\bf r}\sigma}c^i_{{\bf r}\sigma} \rangle$.
The configurations of the orders $m_1$, $m_2$ and $m_3$ are shown in Fig.~\ref{fig:arrange}.
By using Eqs.~(\ref{eq:WMFn})-(\ref{eq:WMF3}), $n^i_{{\bf r}\sigma}$ are written as 
\begin{align}
n^1_{{\bf r}\uparrow}=&\frac{n+(m_1+m_2+m_3)e^{i{\bf Q}\cdot{\bf r}}}{4} \\
n^1_{{\bf r}\downarrow}=&\frac{n+(m_1-m_2-m_3)e^{i{\bf Q}\cdot{\bf r}}}{4} \\
n^2_{{\bf r}\uparrow}=&\frac{n+(-m_1-m_2+m_3)e^{i{\bf Q}\cdot{\bf r}}}{4} \\
n^2_{{\bf r}\downarrow}=&\frac{n+(-m_1+m_2-m_3)e^{i{\bf Q}\cdot{\bf r}}}{4} .
\label{eq:n1n2}
\end{align}
The common part $n/4$ is absorbed in the chemical potential as is the Hartree term originating from the attractive interaction.
Using the above mean fields, we find a mean field Hamiltonian 
\begin{align}
H=& \frac{1}{2}\sum_{\bf k} \Bigl( {\bm \zeta}^{\dagger}_{\bf k} \mathcal{H}_2{\bm \zeta}_{\bf k} +\xi^1_{\bf k}+\xi^1_{{\bf k}+{\bf Q}}+\xi^2_{\bf k}+\xi^2_{{\bf k}+{\bf Q}} \Bigl) \nonumber\\
&+ U|\Delta|^2+ \frac{W}{8}(m^2_1+m^2_2-m^2_3-n^2) .
\label{eq:UWMFH}
\end{align}
Here, we define a $8\times 8$ matrix $\mathcal{H}_2$ and a vector ${\bm \zeta }^{\dagger}_{\bf k}$. 
The matrix $\mathcal{H}_2$ can be diagonalized by using generalized Bogoliubov transformation, such that 
\begin{align}
{\bm \zeta}_{\bf k}=& V_{\bf k} {\bm \beta}_{\bf k}, 
\label{eq:gBog}
\end{align}
where $V_{\bf k}$ is a $8\times 8$ matrix and the elements of a vector ${\bm \beta}_{\bf k}$ are operators of quasi-particles as $\alpha^i_{{\bf k}\sigma}$ in Eqs.~(\ref{eq:Bog1}) and (\ref{eq:Bog2}). 
Substituting this equation into the definitions of the mean fields yields a set of five self-consistent equations.
In the following, as in Sec.~\ref{sec:BI-SF}, we use the equation for the number conservation to define the chemical potential.
Therefore, we solve these five equations self-consistently and obtain the orders $\Delta$, $m_1$, $m_2$, $m_3$ and the chemical potential $\mu$.
In the following, we show numerical results of this mean field treatment.

\subsection{\label{sbusec:integer}Integer-Number Filling}
\begin{figure}
\begin{center}
\includegraphics[width=6cm]{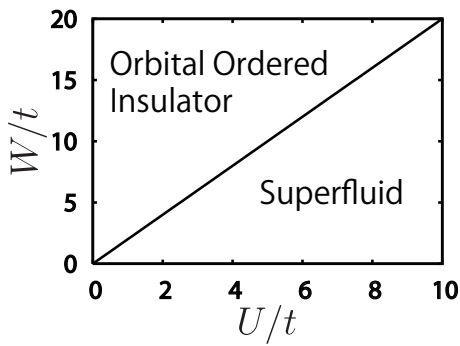} 
\end{center}
\caption{\label{fig:n10PDl} Phase diagram in the plane of repulsive interaction $W$ and attractive interaction $U$ for $\Delta E=0$, $n=2.0$, $t^{\prime}=0.0$ and $T=0$.
The solid line is a first-ordered transition line.
}
\end{figure}
\begin{figure}
\begin{center}
\includegraphics[width=6cm]{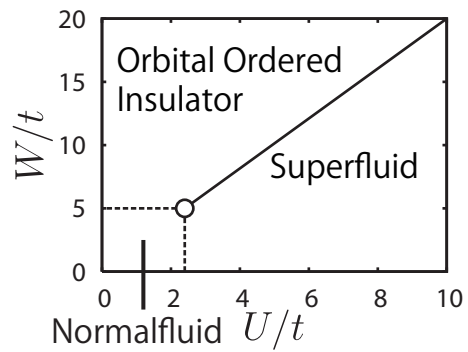} 
\end{center}
\caption{\label{fig:n10PDh} Phase diagram in the plane of repulsive interaction $W$ and attractive interaction $U$ for $\Delta E=0$, $n=2.0$, $t^{\prime}=0.0$ and $T=1$.
The solid line is a first-ordered transition line while the dashed lines show second-ordered transition lines.
The open circle is a bicritical point.
}
\end{figure}
\begin{figure}
\begin{center}
\includegraphics[width=6cm]{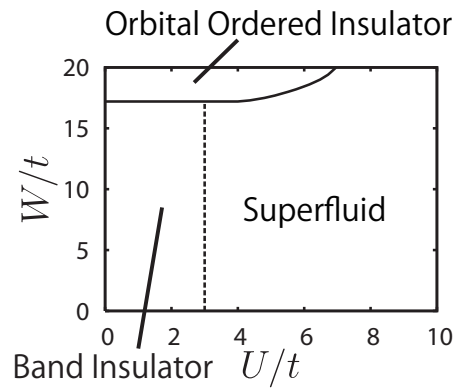} 
\end{center}
\caption{\label{fig:n10PDd} Phase diagram in the plane of repulsive interaction $W$ and attractive interaction $U$ for $\Delta E=9$, $n=2.0$, $t^{\prime}=0.0$ and $T=0$.
The solid line is a first-ordered transition line while the dashed line shows a second-ordered transition line.
}
\end{figure}

In Figs.~\ref{fig:n10PDl}-\ref{fig:n10PDd}, we show phase diagrams for $n=2$ obtained by the above mean field approximation.
At $T=0$ with $\Delta E=0$, two ordered phases are seen in the phase diagram; one is the superfluid (SF) phase and the other is the orbital ordered insulating (OOI) phase.
Strictly speaking, two types for the OOI phase exist.
One is given by the condition with $m_1 \neq 0$ and $m_2=m_3=0$, and the other is given by the condition with $m_2 \neq 0$ and $m_1=m_3=0$.
Both types have the same energy and one of these degenerate two may be realized.
Without loss of generality, we choose the type with $m_1 \neq 0$ as the OOI phase.
When either $T$ or $\Delta E$ differs from zero, however, a non-ordered phase appears in the weak coupling region.
The non-ordered phase at $T=0$ is a normalfluid (N) phase when $|\Delta E|<8$ while it is a band insulating (BI) phase when $|\Delta E|>8$.
Phase transitions between the SF and the N (or BI) phase (the SF/N(BI) transition) are of the second order and the OOI/SF transition is of the first order for any value of $\Delta E$.
On the other hand, the OOI/N(BI) transition is of the second order when $\Delta E=0$ and is of the first order when $\Delta E \neq 0$.
Therefore, the critical point in the phase diagram of Fig.~\ref{fig:n10PDh} for $\Delta E=0$ at $T=1$ is a bicritical point.
When $\Delta E$ is nonzero as in Fig.~\ref{fig:n10PDd}, the OOI/N(BI) transition line is of first order.
In contrast to the phase diagram of Fig.~\ref{fig:n10PDh}, no multicritical point exists in this case. 

Let us focus on the phase diagram for $|\Delta E|>8$.
In this case, the superfluid has two neighboring insulators, the band insulator and the orbital ordered insulator.
As in the case of Sec.~\ref{sec:BI-SF}, the BI phase cannot exist with sufficiently large attractive interaction.
The OOI phase, however, exists, which suppresses the SF phase in the larger $U$ region with the repulsive interaction.
Thus, if the repulsive interaction coexists with the attractive interaction, a new phase, an orbital ordered insulator emerges.
Since this insulator is caused by the repulsive interaction between molecular bosons and exists even in large $U$ regions, it is a candidate for the Mott insulator of \textit{molecular bosons}, though there is symmetry breaking.
In the OOI phase, there exist two types of molecular bosons: one is a molecular boson composed of fermions in the band-1 and the other composed of band-2 fermions.
The emergence of two kinds of molecules is due to the orbital internal degrees of freedom of molecular bosons, which comes from the fermionic degrees of freedom.
Therefore, the OOI phase is a typical example where the orbital internal degrees of freedom of molecular bosons are observed explicitly.
The only example of internal degrees of freedom of bosons so far observed in ultracold atomic gases is the spin degrees of freedom of atomic bosons.
In addition to the spin degrees of freedom of atomic bosons, the orbital degrees of freedom of molecular bosons may also be observed as another internal degrees of freedom of bosons.
Although spin degrees of freedom is often discussed in the context of spinor BECs, the orbital internal degrees of freedom is so far paid less attention.
Since the treatment of the spin degrees of freedom is different from that of the orbital degrees of freedom, more experimental and theoretical works are required to understand the roles of the orbital degrees of freedom on physical properties.

\subsubsection{\label{ssubsec:classification}Classification of Insulators}
Here, we classify insulators consisting both of fermions and of bosons and discuss their relations in order to compare the insulators in our phase diagrams with them.

In Fermi systems, three types of insulators exist, one is a band insulator (BI), another is a Mott insulator (MI) and the last one is an ordered insulator (OI).
We assume that only the OI has symmetry breaking.
In Bose systems, there exist two insulators, one is a boson Mott insulator (BMI) and the other is a bosonic ordered insulator (BOI).

In the superfluid side, it is established that the BCS superfluid and the BEC superfluid are connected each other by a crossover~\cite{crossover1, crossover2, Leggett, NSR}.
In the insulating side, however, it is complicated.
What is established is that a transition should exist between ordered phase and non-ordered phase.
From our phase diagrams, it is likely that an OI and a BOI connect each other by a crossover.
It is reasonable to treat the system in large $U$ region as a Bose system of molecular bosons.
The OOI phase in this region should be classified into a BOI.
On the other hand, in small $U$ region, the system behaves as a Fermi system.
Therefore, the OOI phase in this region should be classified into an OI.
Thus, an OI and a BOI are connected each other by a simple crossover.
About two-component Fermi gases in optical lattices, it was claimed that a BI and a BMI connected each other by a crossover~\cite{BI-SF1,BI-SF2}.
However, so far, it is not clearly settled.
The relation between a BI and an MI is also a fundamental open issue.
The relation between an MI and a BMI is not also established.
In this paper, we established that, because of the orbital internal degrees of freedom, a BOI emerges in a two-component Fermi gas and that an OI and a BOI are connected each other by a simple crossover.
However, the other relations remain as future problems.

\subsubsection{Remaining Questions}
There are two points to be discussed in our phase diagrams.
One is that the OOI/SF transition is of the first order in our phase diagrams while the superfluid-Mott insulator transition in Bose systems is of the second order.
When two local minima in the free energy entangle each other because of quantum fluctuations, these transition may become continuous.
In mean field theories, fluctuations are neglected and thus, the first order transition is favored.
By including fluctuations, however, this transition can be of the second order.

The other point is the critical value of $W$.
In our phase diagrams, the critical value $W_{\rm c}$ is scaled as $W_{\rm c}\sim 2U$ in large $U$ region. 
In Bose systems with repulsive interactions, however, the critical value of the interaction for the superfluid-Mott insulator transition is typically some orders of magnitude smaller than the binding interactions which stabilize atoms and is not scaled as $W_{\rm c}\sim 2U$.
This discrepancy is caused by the overestimate of the kinetic energy of a Cooper pair by the mean field treatment.
Written in real space, the BCS term we considered looks like a pair hopping term to arbitrary distance, namely 
\begin{align}
-\frac{U}{N_{\rm s}} \sum_{\bf kk^{\prime}} c^{i\dagger}_{{\bf k}\uparrow}c^{i\dagger}_{{\bf -k}\downarrow}c^j_{{\bf -k^{\prime}}\downarrow}c^j_{{\bf k^{\prime}}\uparrow} \rightarrow -\frac{U}{N_{\rm s}}\int d{\bm r}d{\bm r^{\prime}}c^{i\dagger}_{{\bf r}\uparrow}c^{i\dagger}_{{\bf r}\downarrow}c^j_{{\bf r^{\prime}}\downarrow}c^j_{{\bf r^{\prime}}\uparrow} .
\end{align}
The kinetic energy of a Cooper pair is scaled with $U$ in the mean field approximation.
Therefore, it is reasonable that the critical value $W_{\rm c}$ scales with $U$ in our approximation.
Let us consider the region where $U$ is large and $W=0$ by perturbation theories.
In this region, however, the kinetic energy of a Cooper pair is estimated at about $t^2/U$.
Thus, the critical value of the repulsive interaction for the OOI/SF transition should be scaled as $W_{\rm c}\sim t^2/U$, which is consistent with the difference between $W_{\rm c}$ and $U$ by orders of magnitude in simple Bose systems.
We believe that more sophisticated treatment beyond the mean field approximations would not face these problems, though most essential results are described in our mean field treatment.

This scaling indicates that, in large $U$ regions, the critical value of $W$ is relatively small.
These conditions are realized when detuning from resonant points is larger.
This means that such regions are favorable in experiments because losses of atoms from traps are suppressed.

\subsection{\label{subsec:coexistence}Noninteger Filling: Coexisting Phase}
\begin{figure}
\begin{center}
\begin{tabular}{c}
\includegraphics[width=6cm]{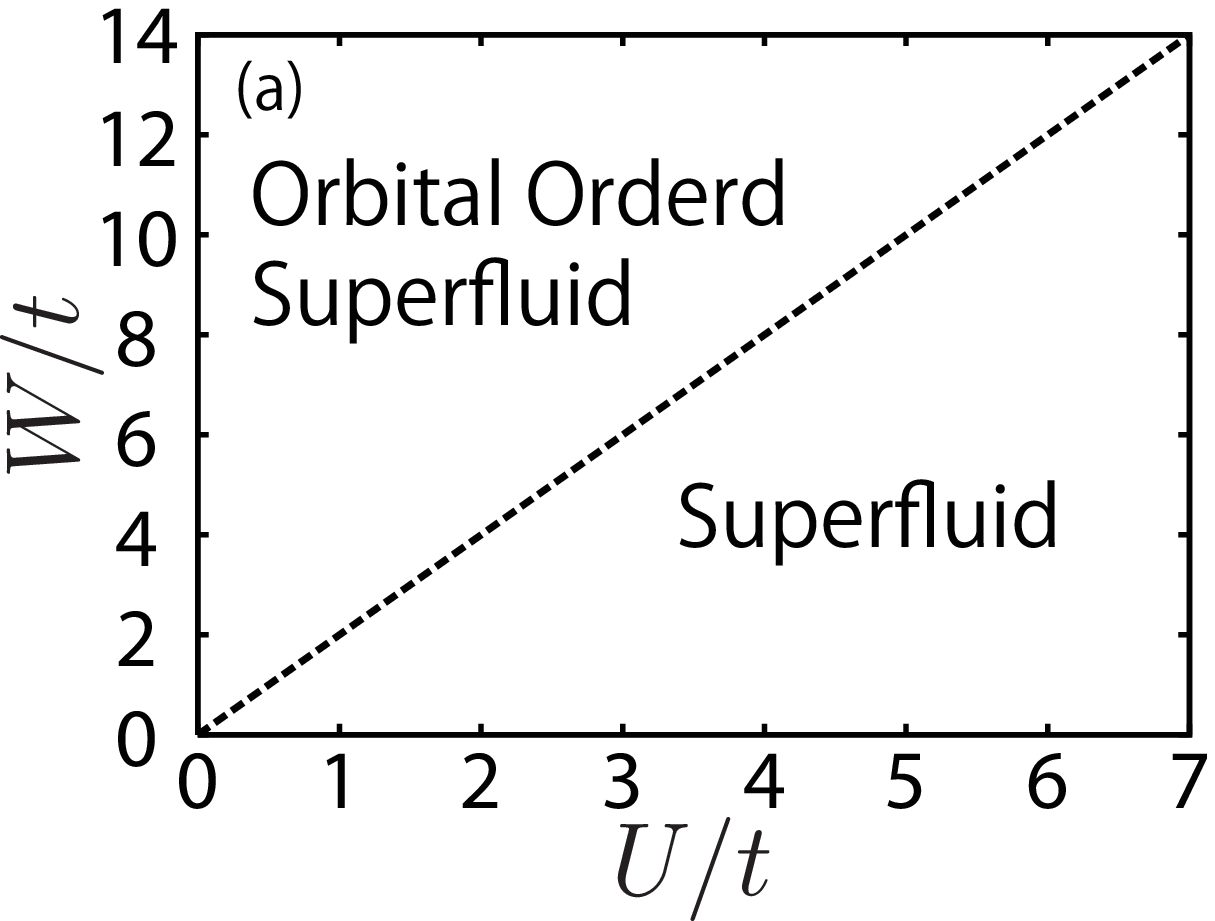} \\
\includegraphics[width=6cm]{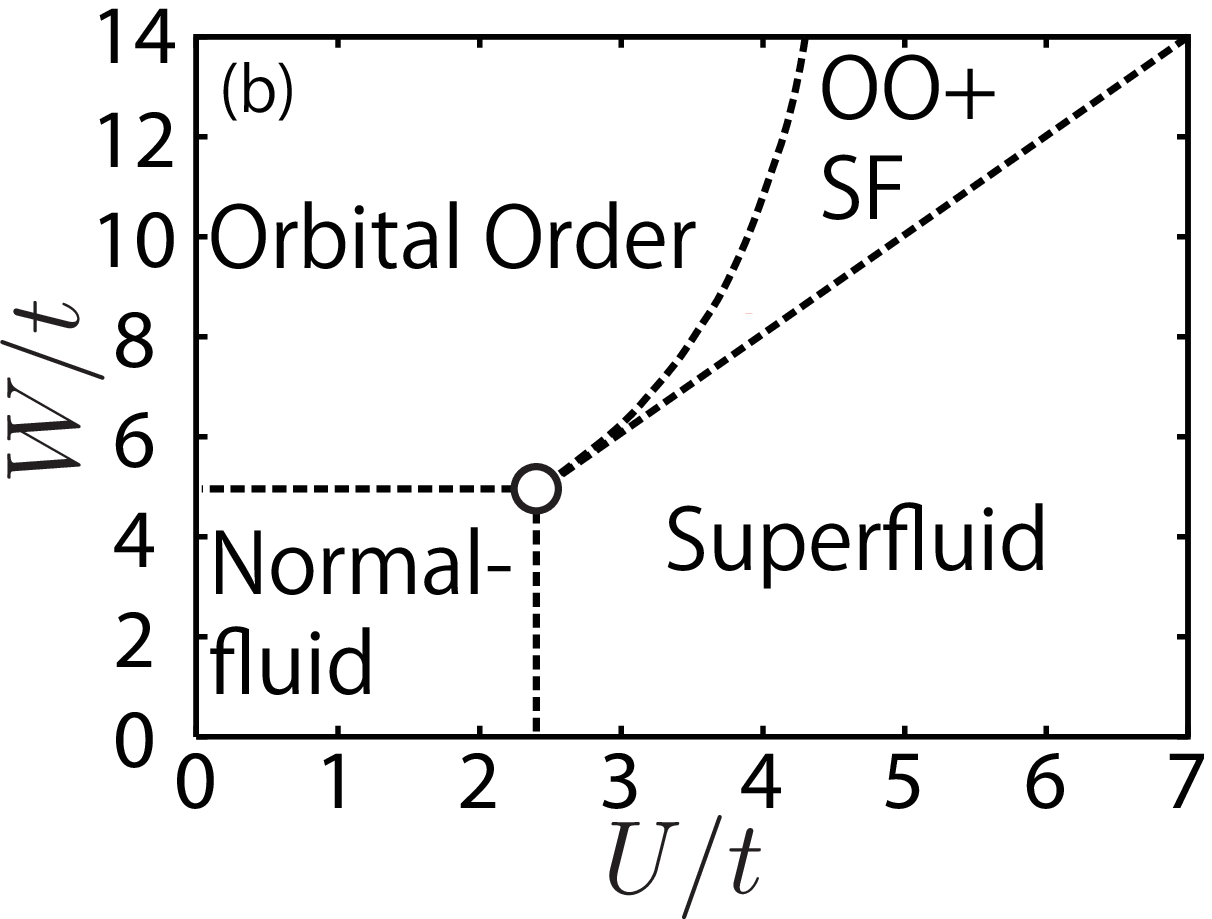} \\
\end{tabular}
\caption{\label{fig:n09PD} Phase diagrams in the plane of repulsive interaction $W$ and attractive interaction $U$ at $\Delta E=0$, $n=1.8$ and $t^{\prime}=0.0$.
The upper panel (a) is at $T=0$ and the lower panel (b) is at $T=1$.
In both phase diagrams, coexisting phases appear.
The dashed lines show second-ordered transition lines.
The open circle in the panel (b) is a tetracritical point.
}
\end{center}
\end{figure}
\begin{figure}
\begin{center}
\begin{tabular}{c}
\includegraphics[width=6cm]{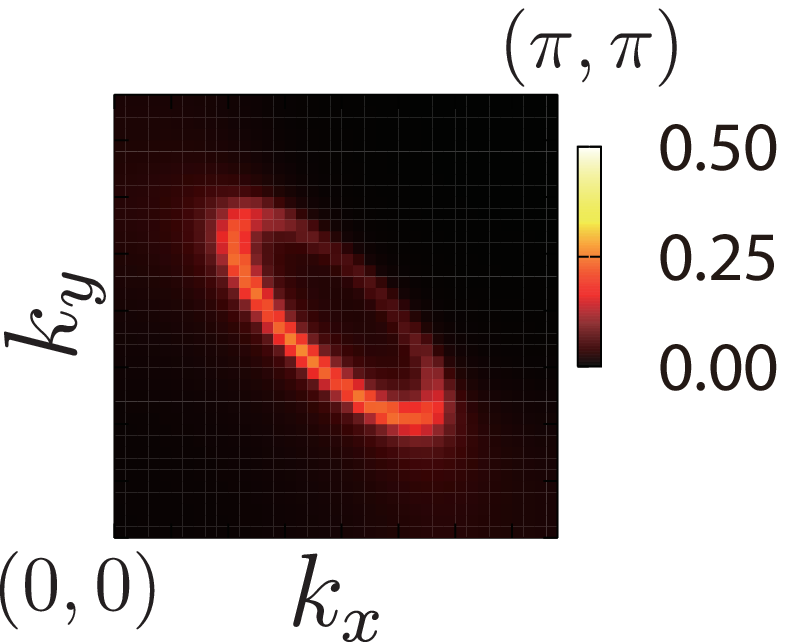} \\
\includegraphics[width=6cm]{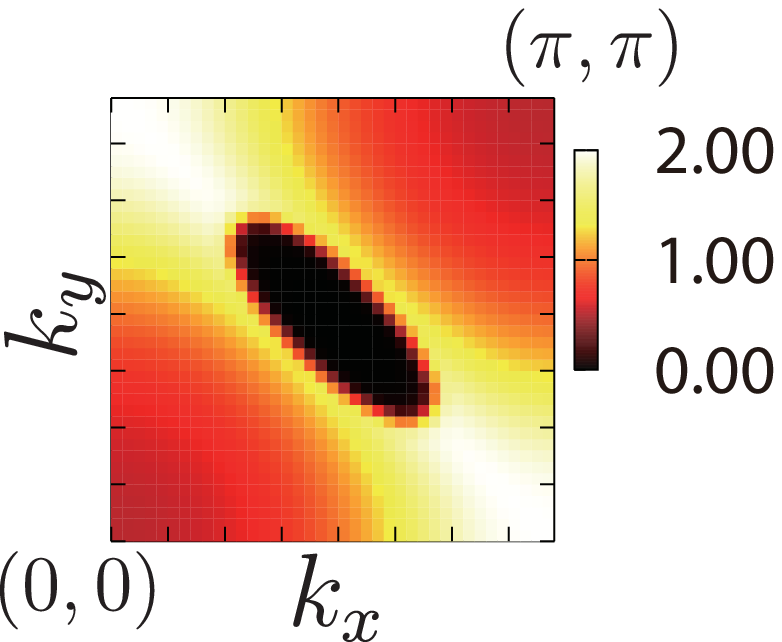} 
\end{tabular}
\caption{\label{fig:momentum} (color online) SF and OO amplitudes in momentum space at $\Delta E=0$, $t^{\prime}=0.2$, $n=1.8$, $W=4$ and $U=0.5$.
The upper panel is the SF amplitude while the lower panel is the OO amplitudes.
}
\end{center}
\end{figure}

Density of particles is also easily tuned in ultracold atomic gas systems.
We now show phase diagrams when $n$ deviates from 2 for $\Delta E=0$ in Fig.~\ref{fig:n09PD}.
Besides the SF and the orbital ordered (OO) phase, a new phase appears where the SF order coexists with the orbital order (the OO+SF phase).
Both of the OO/OO+SF and the OO+SF/SF transitions are of the second order in contrast to the case of $n=2$ where the OO/SF transition is of the first order.
In the phase diagram at $T=1$, a tetracritical point is found while it is replaced with the bicritical point when $n=2$.
However, at absolute zero temperature, the critical point is located at $(U, W)= (0, 0)$ and the OO phase without the superfluid order vanishes.

In the coexisting phase, the amplitude of the superfluidity $n_{\rm sf}({\bm k})$ and that of the orbital order $n_{\rm oo}({\bm k})$ is well separated in momentum space.
The definition of the amplitudes $n_{\rm sf}({\bm k})$ and $n_{\rm oo}({\bm k})$ are 
\begin{align}
n_{\rm sf}({\bm k})=& \langle c^1_{-{\bf k}\downarrow}c^1_{{\bf k}\uparrow} \rangle , \\
n_{\rm oo}({\bm k})=& \sum_{\sigma} \langle c^{1\dagger}_{{\bf k}+{\bf Q}\sigma}c^1_{{\bf k}\sigma} - c^{2\dagger}_{{\bf k}+{\bf Q}\sigma}c^2_{{\bf k}\sigma}\rangle .
\end{align}
In Fig.~\ref{fig:momentum}, we show both the amplitude of the orbital order and that of the superfluidity in momentum space.
Similarly to the Fermi surface observed in underdoped regions of the high-$T_{\rm c}$ superconducting cuprates~\cite{arc1,arc2,arc3}, the amplitude of the superfluidity is large around $(\pi/2, \pi/2)$.
Although this system is completely different from the cuprates and the similarity is superficial in many respects, these differentiations suggest the existence of underlying common physics~\cite{Onoda, Sakai}.
Ultracold atomic gases are ideal systems to investigate how robust this differentiation in momentum space exists when we control the interactions.

\section{\label{sec:sum}Summary}
We have studied two different systems which can be designed by two-component Fermi gases.
One is the case with attractive interaction between the opposite components, which is similar to the situation in the experiment by Zwierlein {\it et al.}~\cite{experiment-MIT}.
The other is the case both with attractive and repulsive interactions, which is realized by utilizing more than one Feshbach resonances.

In the first system, there should be a band insulator-BEC superfluid transition when one changes the interaction strength.
This is in contrast with the claim by Zwierlein {\it et al.} although they claim the transition between the superfluid and Mott insulator of \textit{molecular bosons}.
The origin of this transition is the energy gain coming from the BCS channel (or pair formation in higher bands) as we discussed in Sec.~\ref{sec:BI-SF}.
This band insulator-BEC superfluid transition is characterized by the following features: 

\noindent(1) The excitation gap $\Delta_{\rm gap}$ is not scaled by the superfluid order parameter $\Delta$ in the vicinity of the transition in contrast to the conventional superfluid-normalfluid transition.
Indeed, at the transition point, $\Delta_{\rm gap}$ is even nonzero while $\Delta$ vanishes.
This is consistent with the previous results in Refs.~\cite{BI-SF1,BI-SF5}.
The excitation spectra may be observed by the photoemission spectroscopy which is recently developed by Stewart {\it et al.}~\cite{PE}.

\noindent(2) The transition is that between a Bose-Einstein condensation of molecular bosons composed of two fermions and a band insulator of fermions.
This is confirmed by the large binding energy of Cooper pairs near the transition point.

\noindent(3) With decreasing temperatures, a reentrant transition into the non-ordered phase appears.

In the second system, in the interaction region where treating a molecular boson as a minimum unit is not justified, we have found a new insulating phase, an orbital ordered insulator (OOI).
Differing from a band insulator, the OOI has symmetry breaking.
The symmetry breaking is an outstanding example of effects of internal degrees of freedom of molecular bosons which is observable.
The OOI is also one candidate for the Mott insulator of {\it molecular bosons}, though it has symmetry breaking in the present mean-field theory and it can be replaced with a genuine Mott insulator without any symmetry breaking if more sophisticated treatment is employed for lattices with a geometrical frustration effect.
Besides these results, we find in general a coexisting phase in the second system when density of particles $n$ deviates from 2.
By doping holes or particles into the orbital ordered insulator, the phase where the superfluid order coexists with the orbital order emerges.
The coexistence shows up as a differentiation in momentum space.

These properties obtained in two systems are characteristic for the intermediate regions where the attractive interaction energy is comparable to the band gap energy or the repulsive interaction, where multi-band effects are crucial.

\begin{acknowledgments}
One of the authors (R. W.) thanks Y. Yamaji and H. Hirayama for useful discussions.
This work is supported by Grants-in-Aid for Scientific Research on Priority Areas under the grant number 17071003 from MEXT, Japan.
\end{acknowledgments}

\end{document}